\documentclass{aa}  

\newcommand{\msun}{{$M_\odot$}}
\newcommand{\miriad}{{{MIRIAD}}}
\newcommand{\HI}{{\sc H\,i}}
\usepackage{hyperref}
\usepackage{courier}
\usepackage{graphicx}
\usepackage{grffile} 
\usepackage{txfonts}
\usepackage{natbib}
\usepackage[flushleft]{threeparttable}

\usepackage{color}
\usepackage{amstext}

\begin{document}

   \title{LOFAR discovery of a 700-kpc remnant radio galaxy at low redshift}

   \author{M. Brienza
          \inst{1,2}\fnmsep\thanks{brienza@astron.nl}
          \and
           L. Godfrey\inst{1}
          \and
          R. Morganti\inst{1,2}
          \and
          N. Vilchez\inst{1}
          \and
          N. Maddox\inst{1}
          \and
          M. Murgia\inst{3}
                  \and
          E. Orru\inst{1,4}
          \and
          A. Shulevski\inst{1}                    
                  \and
                  P. N. Best\inst{5}
                  \and
                  M. Br\"uggen\inst{6}
                  \and
                  J.  J. Harwood\inst{1}
                  \and
                  M. Jamrozy\inst{7}
                  \and
                  M. J. Jarvis\inst{8}
                  \and
                  E. K. Mahony\inst{1}
          \and
          J. McKean\inst{1,2}
          \and
                  H.J.A. R\"ottgering\inst{9}
                  }

   \institute{ASTRON, the Netherlands Institute for Radio Astronomy, Postbus 2, 7990 AA, Dwingeloo, The Netherlands
            \and
             Kapteyn Astronomical Institute, Rijksuniversiteit Groningen, Landleven 12, 9747 AD Groningen, The Netherlands
                        \and
                         INAF – Osservatorio Astronomico di Cagliari, Loc. Poggio dei Pini, Strada 54, 09012 Capoterra (CA), Italy             
             \and
             Department of Astrophysics/IMAPP, Radboud University, P.O. Box 9010, 6500 GL Nijmegen, The Netherlands
             \and
             SUPA, Institute for Astronomy, Royal Observatory, Blackford Hill, Edinburgh, EH9 3HJ, United Kingdom
                        \and
                        Hamburger Sternwarte, University of Hamburg, Gojenbergsweg 112, 21029 Hamburg, Germany               
                        \and
                        Obserwatorium Astronomiczne, Uniwersytet Jagielloński, ul Orla 171, 30-244, Kraków, Poland                     
                        \and
                        Astrophysics, University of Oxford, Keble Road, Oxford, OX1 3RH, UK
                        \and
                        Leiden Observatory, Leiden University, Niels Bohrweg 2, 2333 CA Leiden, The Netherlands}

   \date{Received 16 June 2015/ Accepted 17 August 2015 }

\abstract
{Remnant radio galaxies represent the final dying phase of radio galaxy evolution in which the jets are no longer active. Remnants are rare in flux-limited samples, comprising at most a few percent. As a result of their rarity and because they are difficult to identify, this dying phase remains poorly understood and the luminosity evolution is largely unconstrained.}
{Here we present the discovery and detailed analysis of a large (700 kpc) remnant radio galaxy with a
low surface brightness that has been identified in LOFAR images at 150 MHz.}
{By combining LOFAR data with new follow-up Westerbork observations and archival data at higher frequencies, we investigated the source morphology and spectral properties from 116 to 4850 MHz. By modelling the radio spectrum, we probed characteristic timescales of the radio activity.}
{The source has a relatively smooth, diffuse, amorphous appearance together with a very weak central compact core that is associated with the host galaxy located at z=0.051. From our ageing and morphological analysis it is clear that the nuclear engine is currently switched off or, at most, active at a very low power state. We find that the source has remained visible in the remnant phase for about 60 Myr, significantly longer than its active phase of 15 Myr, despite being located outside a cluster. The host galaxy is currently interacting with another galaxy located at a projected separation of 15 kpc and a radial velocity offset of $\sim 300 \ \rm km \ s^{-1}$. This interaction may have played a role in the triggering and/or shut-down of the radio jets. }
{The spectral shape of this remnant radio galaxy differs from most of the previously identified remnant sources, which show steep or curved spectra at low to intermediate frequencies. Our results demonstrate that remnant radio galaxies can show a wide range of evolutionary paths and spectral properties. In light of this finding and in preparation for new-generation deep low-frequency surveys, we discuss the selection criteria to be used to select representative samples of these sources.}


   \maketitle


\section{Introduction}

Radio-loud active galactic nuclei (AGN) can have a substantial energetic impact on both their host galaxy and the surrounding intergalactic medium \citep{bruggen2002, croton2006, morganti2013}. The jets of relativistic particles produced by the AGN can reach distances of up to a few Mpc from the host galaxy \citep{willis1974, machalski2008}. The active phase of the radio AGN can last several tens of Myr, after which the nuclear activity stops and the source starts to fade away \citep{parma1999}. In this fading period, often termed the remnant or dying phase, the core and the jets disappear and only the lobes remain visible, radiating their energy away \citep{slee2001}. When an accurate knowledge of the radio spectral characteristics is available, remnant radio galaxies provide a timing of the on-off cycle of the radio activity, which constrains the evolutionary history of the radio source \citep{kardashev1962, pacholczyc1970, jaffe1973}. The fate of the AGN remnant radio plasma and the physical processes driving its evolution are widely relevant to studies of radio galaxy feedback. Moreover, a better knowledge of these objects could also give new insights into the formation and evolution of cluster AGN relics and radio phoenices \citep{slee2001, ensslin2002, vanweeren2009, degasperin2015}.

Statistical studies based on number densities of radio sources show that radio-loud AGN are active for a short interval of time with respect to the entire life of the host galaxy. The duty cycle of the radio-loud AGN strongly depends on the power of the radio source \citep{best2005, shabala2008}. Therefore, we would expect to observe many sources showing signatures of remnant emission from past radio-loud AGN activity. But detections of such objects have always been elusive and rare. \cite{giovannini1988} identified only a few percent of the radio sources in the B2 and 3C samples as remnant radio galaxies. The reason for this rarity is most likely connected to the rapid luminosity evolution of the fading plasma \citep{kaiser2002}. Synchrotron cooling and inverse Compton scattering of cosmic microwave background photons cause a steepening of the spectrum at high frequency
that renders the source undetectable. In addition, adiabatic expansion of the lobes contributes further energy losses in an energy-independent way, decreasing the flux densities across the whole spectrum. Although still poorly understood, the timescales of  these processes depend on the physical conditions within the lobes and the ambient medium as the jets switch off. For example, if the lobes remain strongly overpressured at the end of the active phase, adiabatic losses can be significant, unlike the case where the lobes have already reached pressure equilibrium with the surrounding gas. Moreover, powerful sources with high magnetic field values are subject to much higher radiative losses.

Systematic studies of remnant radio galaxies as a population have always been hampered by the small number of sources available and by non-uniform selection effects that limit the study of their characteristics and evolution. The known objects cover a variety of morphologies, spectral characteristics, and physical conditions. Dying radio galaxies typically show relaxed morphologies without compact components and curved steep spectra \citep{jamrozy2004, giacintucci2007, parma2007, murgia2011} even at frequencies $\rm \leq1$ GHz. In some cases, the fossil lobes look like bubbles of emitting plasma with a very irregular morphology \citep[e.g.][]{giovannini1991, harris1993, degasperin2014}. Since radio jets can be episodic in nature, remnant plasma can also be associated with a pair of new-born jets \citep {clarke1991, jones2001, saikia2009}. These sources are often termed restarted radio sources. The most common example of these are the double double radio galaxies \citep[DDRG,][]{schoenmakers2000, nandi2012} that posess a pair of diffuse, steep-spectrum lobes which is well detached from the renewed radio activity. Other evidence of this occurrence is seen in the diffuse, extended emission surrounding some GHz-peaked spectrum and compact steep spectrum sources \citep{baum1990, stanghellini2005, shulevski2012}. We note that most of the remnant sources have been found in massive clusters in which the dense intergalactic medium prevents strong expansion losses \citep{murgia2011}. However, a few examples in low-density environments have also been identified \citep{jamrozy2004, degasperin2014, hurleywalker2015}.

To date, a unified physical framework in which to interpret the variety of individual studies mentioned above has not been provided.  
To constrain models describing the final stages of the life of radio galaxies, we must take a statistical approach \citep{kaiser2002} and, therefore, must expand the search for remnant radio galaxies. Greater statistics of their characteristics will enable an investigation of the physical processes driving the remnant plasma evolution as a function of source power and environment. With this aim in mind, we plan to exploit upcoming deep and wide field radio surveys to build up complete samples of such sources. These new facilities are designed to significantly enhance the detection probability of these rare low-surface brightness objects. In particular, remnant emission is expected to be brighter at low radio frequencies and, therefore, instruments such as the Low-frequency Array \citep[LOFAR,][]{vanhaarlem2013} and the Murchison Widefield Array \citep[MWA,][]{lonsdale2009} will play a key role in the systematic search for them. However, it is difficult to identify
remnant radio galaxies, and simple selection criteria based on integrated spectra alone can result in a biased representation of the remnant population. For this reason, it is important to address the issues of remnant selection and associated selection effects.

A complementary approach to remnant radio galaxy selection is radio morphology.  We have preliminarily explored the available LOFAR fields to search for large radio sources with a low surface ratio without compact features, which may be indicative of remnant radio galaxies. Here we present one 700 kpc remnant radio galaxy at coordinates J2000.0 RA 18h 28m 20.4s Dec +49d 14m 43s (we
call this blob1) that was discovered as part of this search. blob1 was first identified on purely morphological arguments based on its unusually large angular extension with amorphous shape, low surface-brightness, and lack of compact features. We investigate the physics of blob1 in comparison with already known active and remnant radio galaxies with the  aim to reconstruct the evolution history of the radio source from its birth to its death. This is also the approach that we intend to use for future sources. Furthermore, we discuss the selection criteria to be used for extracting complete samples of this population of sources out of the upcoming deep low-frequency surveys. Such studies are an important first step to help guide, plan, and interpret a broader search for remnant radio galaxies with LOFAR.

The paper is organized as follows: in Sect. \ref{data} we describe the data and the data reduction procedures; in Sect. \ref{results} we present the source properties resulting from the data analysis; in Sect. \ref{discussion} we investigate the history of the source, addressing the question of how this object fits in the current scenario of active and remnant radio galaxies, and what its evolution history is. Finally, in Sect. \ref{implicationsforsourceselection} we discuss the implications of this study for selecting remnant radio galaxies in upcoming surveys. The cosmology adopted throughout the paper assumes a flat universe and the following parameters: $\rm H_{0}= 70\  km \ s^{-1}Mpc^{-1}$, $\Omega_{\Lambda}=0.7,
\text{and }  \Omega_{M}=0.3$. The spectral index, $\alpha$, is defined using the convention $F_\nu=\nu^{-\alpha}$.


\section{Data}
\label{data}

\subsection{LOFAR observations}
\label{lofar}

blob1 has been detected with the High Band Antenna (HBA) within the field of view centred on the LOFAR primary calibrator 3C380 at a distance of about 30\arcmin \  from the phase centre. The observations took place on March 14, 2014 as part of cycle 1 for a total integration time of 8 h. The sampling time was set to 1 second, and four polarization products (XX, YY, XY, and YX) were recorded. The total bandwidth, equal to 44.3 MHz, was divided by default into 227 sub-bands of 195.3 kHz with 64 frequency channels. Only the Dutch LOFAR stations were included in the observation, giving a maximum baseline of 100 km. We summarize the details of the observational configuration in Table ~\ref{tab:Lofarconfiguration}. 

The data were pre-processed by the observatory pipeline \citep{heald2010} as follows. Each sub-band was automatically flagged for RFI using the AOFlagger \citep{offringa2012} and was averaged in time to 5 seconds per sample and in frequency by a factor of 16, giving four channels per sub-band in the output data. The calibrator data were used to derive antenna gains, solving for a diagonal Jones matrix using the black board self-calibration tool \citep[BBS,][]{pandey2009}, which takes into account the variation of the LOFAR station beams
with time and frequency. The source 3C380 was used as flux density calibrator, and the flux scale was set following \cite{scaife2012}. 

The amplitude-corrected visibilities were merged in chunks of 20 sub-bands (3.9 MHz) and the channels were averaged to 1 per sub-band. Each frequency chunk was imaged separately using the \texttt{awimager} software  \citep{tasse2013} and fixing the field of view to $\rm 5^{\circ}\times 5^{\circ}$. This imager makes use of the aw-projection algorithm to model the primary beam variations with time and frequency across the field of view. Moreover, it corrects for the effects of non-coplanar baselines when imaging a large field of view.

Phase self-calibration was performed iteratively, adding progressively longer baselines to achieve the final resolution in three steps (Vilchez et al. in prep.). As a starting point, only core stations were used so that the phase coherence was preserved and because they are set to the same clock. The initial phase-calibration model was derived from the VLSS catalogue and contains all sources located in the field of view with flux density higher than 4 Jy. For each VLSS source, counterparts in the WENSS and NVSS catalogues were searched for and associated with these sources to provide spectral index information (see Sect. 6.4 in the LOFAR cookbook\footnote{\url{https://www.astron.nl/radio-observatory/lofar/lofar-imaging-cookbook}} for more details). We here took into account a first-order effect of the ionosphere in the direction of the dominant source, but did not perform any direction-dependent calibration.

The images were produced using a robust weighting of -0.3 and have a final beam size of $\rm 59\times37\arcsec $. The final image was obtained by combining all images in the image plane,
which resulted in a central frequency of 137 MHz and a noise of 3.5 mJy beam$^{-1}$. For the spectral analysis presented in Sect. \ref{spectralpropertiesandages}, we used two different sets of 20 sub-bands centred at 116 and 155 MHz, respectively. The integrated flux densities of the source extracted from the maps are listed in Table~\ref{tab:fluxes}. The errors on the measurements were computed taking into account different contributions: i) 2\% due to the uncertainty in the flux scale \citep{scaife2012}, ii) 10\% due to the uncertainty in the global beam model \citep{vanweeren2014}, iii) 15\% due to variations in the measured flux density when using apertures of different sizes, iv) $\sim$ 10\% due to the statistical error (thermal noise) measured around the source following \cite{klein2003}.

\begin{table}[h]

        \small
 \caption{Details of the LOFAR observations.}
        \centering
                \begin{tabular}{l l}
                \hline
                \hline
                Obs. ID &  L206016 \\
                Date & 14-March-2014 \\
                Phase centre 3c380 (J2000.0)& 18h 29m 31.8s, +48d 44m 46s\\
                Target coordinates (J2000.0) & 18h 28m 20.5s, +49d 14m 20s \\
                Total no. of stations used & 60 \\              
                No. of remote stations & 14 \\
                Integration time & 8 hours \\
                Sampling time & 1 second \\             
            Polarization & Full Stokes \\
                Frequency range & 115-162 MHz \\
            Band width  & 44.3 MHz \\
            Synthesized beam &  \\
            (robust = -0.3) &$59 \times 37$\arcsec \ PA = -45$^\circ$\\
            RMS (robust = -0.3) & 3.5 $\rm mJy \ beam^{-1}$\\
                \hline
                \hline  
                \end{tabular}
   
        \label{tab:Lofarconfiguration}
\end{table}

\subsection{WSRT observations}

Because blob1 is not detected in the NRAO VLA Sky Survey \citep[NVSS,][]{condon1998}, we carried out deeper 1.4-GHz observations using the Westerbork Synthesis Radio Telescope (WSRT) on 12 and 24 August 2014. We also used the WSRT at high resolution to observe the field at 4.9 GHz on 29 March 2015 to determine whether there are any compact components. The observations and data reduction are described below, and the setup is also summarized in Table \ref{tab:wsrtconfiguration}.

\bigskip
\bigskip

\noindent\textit{Observations at 1.4 GHz} 

\noindent In the first run, we observed for 12 h using the standard continuum setup of eight sub-bands of 20 MHz with 64 channels in each sub-band. Of these, four had technical problems and were discarded. In the second run, we observed for 10 h with six sub-bands of 20 MHz with 64 channels for each sub-band. An extra sub-band of 20 MHz and 256 channels was dedicated to explore possible \HI\ spectral lines and was centred on 1350 MHz to include HI at the redshift of the two candidate host galaxies (see Sect.\ 3.2). Primary calibrators (3C286 and 3C147) were observed at the beginning and end of each run. 

The data were calibrated and reduced using the \miriad\footnote{\url{http://www.atnf.csiro.au/computing/software/miriad/}} software package \citep{Sault1995}. Using standard cross calibration and self-calibration, the image is dominated by off-axis phase errors associated with a strong off-axis source (3C380), situated at $\sim 30$\arcmin \ from the field centre. These off-axis errors are most likely due to pointing or tracking errors and/or frequency-dependent errors in the primary model that are due to standing waves in the dishes of the WSRT and cannot be corrected with standard direction-independent calibration. Therefore, we used the technique commonly known as peeling, following the standard steps\footnote{\url{http://www.astron.nl/~oosterlo/peeling.pdf}}
\citep{noordam2004} to correct for off-axis errors in the image. The procedure involves the subtraction of the central sources in the field and self-calibration on individual problematic off-axis sources (in this case, one source, 3C380), iteratively producing antenna-based corrections and a source model for different directions. The source model for 3C380 obtained in this way, including the additional off-axis corrections, was then subtracted from the original visibility data set. The peeling procedure was run separately
on each 20 MHz sub-band.

The final image of blob1 was obtained in two ways. In the first case, we separately imaged each sub-band (i.e.\ 20 MHz) and combined the final images in the image plane. With the second approach, we produced an image by combining all the data in the $uv$ plane. The two different strategies resulted in very similar maps. In both cases, we used a combination of robust weighting 0.4 and a 25\arcsec\  taper to achieve a low side-lobe level and a final resolution that allows imaging the low surface brightness emission expected for blob1. The final beam size is $\rm 30\arcsec \times 27\arcsec$ (PA$ = -3^\circ$) and the noise of the image is 80 $\mu$Jy beam$^{-1}$. The image is presented in Fig.\ref{fig:maps}. To image and locate compact components, we also produced an image with uniform weighting, reaching a resolution of $\rm 15\arcsec \times 9\arcsec$ (PA$ = -4^\circ$) and noise equal to 50 $\mu$Jy beam$^{-1}$. The integrated flux density used for the spectral analysis was derived using only data from the first observing run, which have better quality. This is reported in Table~\ref{tab:fluxes}. The error on the measurements was computed taking into account different contributions: i) 2\% due to the uncertainty in the flux scale, ii) 15\% due to variations in the measured flux density when using apertures of different sizes, iii) <1\% due to the statistical error (thermal noise) measured around the source following \cite{klein2003}, and iv) 20\% due to variations in the measured flux density when using distinct cleaning strategies.

No \HI\ emission was detected. However, because subtracting 3C380
was difficult, the \HI\ cube has a high noise (about 0.8~mJy beam$^{-1}$ per channel with a channel width equal to 16.5 km s$^{-1}$), making the search for \HI\ not sensitive enough for useful constraints. 
The corresponding upper limit of \HI\ mass is derived following \cite{emonts2010}. At a redshift of $z = 0.05$ (see Sect.\ref{opticalidentifcation})  and assuming a potential $3\sigma$ detection smoothed across a velocity range of 200 $\rm km \ s^{-1}$ (typical for galaxies detected in \HI\ ), we derived an upper limit of $\rm M_{HI} < 1.5 \times 10^9$ \rm\msun.

\bigskip

\noindent\textit{Observations at 4.9 GHz} 

\noindent  The field was observed for 11 h in a standard continuum configuration (see Table \ref{tab:wsrtconfiguration}). The array configuration was not optimized for detecting large-scale structure emission but for identifying compact components at high
resolution. A standard data reduction was performed using \miriad. The final image, obtained with uniform weighting, has a synthesized beam of $ 4.03 \times  2.94$\arcsec (PA 0.5 deg) and a noise level of 85 $\mu$Jy.

\begin{small}

\begin{table}[h]
        \centering
        \caption{ List of flux densities and respective errors. }
        \small
                        
                \begin{tabular}{*4c}
                \hline
                \hline
                Telescope & Frequency    & Flux density & Error \\
                &(MHz)  & (mJy) & (mJy) \\
                \hline
                LOFAR &  116 &     1400 & 300\\
                LOFAR & 155  &   1200 & 250 \\
                WENSS & 327  &    800 & 150\\
                WSRT & 1400 &   215 & 45 \\
                GB6 & 4850 &  < 30&-\\
                \hline
                \hline  
                \end{tabular}
                \label{tab:fluxes}
\end{table}
\end{small}

\begin{table}[h]
        \centering
        \small
         \caption{Details of the WSRT observations.}
        \centering
                \begin{tabular}{l l}
                \hline
                \hline
            RA, Dec (J2000.0)  & 18h 27m 01.9s  +49d 12h 25.9s\\
        \hline          
                \sl{1.4 GHz - continuum}\\
                \hline          
                Date & 12, 24-Aug-2014\\
                Integration time & 22 hours \\
                Frequency  & 1400 MHz \\
            Band width  & $4 \times 20$ MHz$^*$ + $6 \times 20$ MHz$^{**}$ \\
            Synthesized beam  &   \\           
                (robust=0.4)    & $35 \times  26$\arcsec \ PA = $-10.4^\circ$ \\
                RMS (robust=0.4)& 0.08  $\rm mJy \ beam^{-1}$\\ 
                Synthesized beam  &   \\               
                (uniform weighting) & $15 \times  9$\arcsec \ PA = $-4^\circ$\\
                RMS (uniform weighting) & 0.05  $\rm mJy \ beam^{-1}$\\                 
                \hline
                \sl{1.4 GHz - \HI} \\
                \hline              
                Central frequency  & 1350 MHz \\
            Band width  & 20 MHz \\          
            Number of channels & 256 \\
            Velocity resolution & 16.5 $\rm km \ s^{-1}$\\ 
            \hline              
                \sl{4.9 GHz} \\
                \hline
                Date & 29 March 2015 \\
            Integration time & 11 hours \\
            Frequency  & 4900 MHz \\
            Band width  & $8 \times 20$ MHz \\
            Synthesized beam  & $4\times  3$\arcsec\ PA = $0.5^\circ$ \\
                (uniform weighting) \\
                RMS (uniform weighting) & 0.085         $\rm mJy \ beam^{-1}$\\      \hline
                \hline  
                \end{tabular}
                \begin{tablenotes}
      \small
      \item \textbf{Notes.} $^*$usable bands in the first observation; $^{**}$ usable bands in the second observation.
    \end{tablenotes}
   
        \label{tab:wsrtconfiguration}
\end{table}

\subsection{Archival radio data}
 
We searched images of the relevant area from several radio surveys (i.e. VLA Low-Frequency Sky Survey, VLSS, \cite{cohen2007}; Faint Images of the Radio Sky at Twenty-cm \cite{becker1995} and NVSS \cite{condon1998}), and we only identified the source in the Westerbork Northern Sky Survey \citep[WENSS,][]{rengelink1997} at 327 MHz (see Fig.~\ref{fig:maps}). However, we also obtained a useful upper limit at 4.8 GHz from the Green Bank 6 cm radio survey \citep[GB6,][]{condon1994} map, which was used in modelling the source spectrum (see Sect. \ref{spectralpropertiesandages}). The upper limit was extracted from the GB6 map and fixed at $3\sigma,$ where $\sigma$ was determined from a set of flux density measurements surrounding the source location using a box with the same size as the source. The parameters of these maps are listed in Table \ref{tab:fluxes}.

\section{Results}
\label{results}

\subsection{Morphology}
\label{morphology}

\begin{figure}[h!]
\centering
\includegraphics[width=0.4\textwidth]{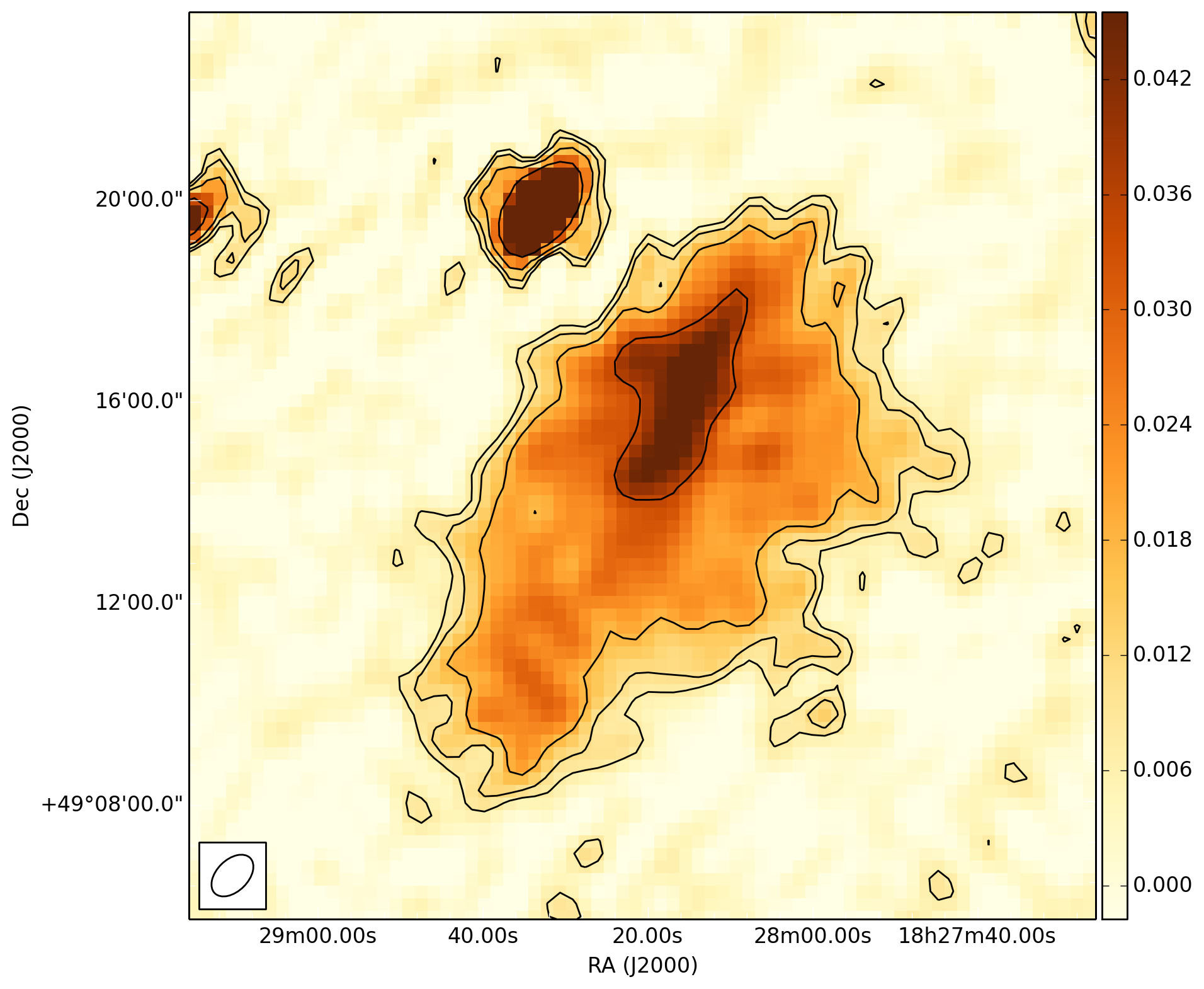}

\centering
\includegraphics[width=0.4\textwidth]{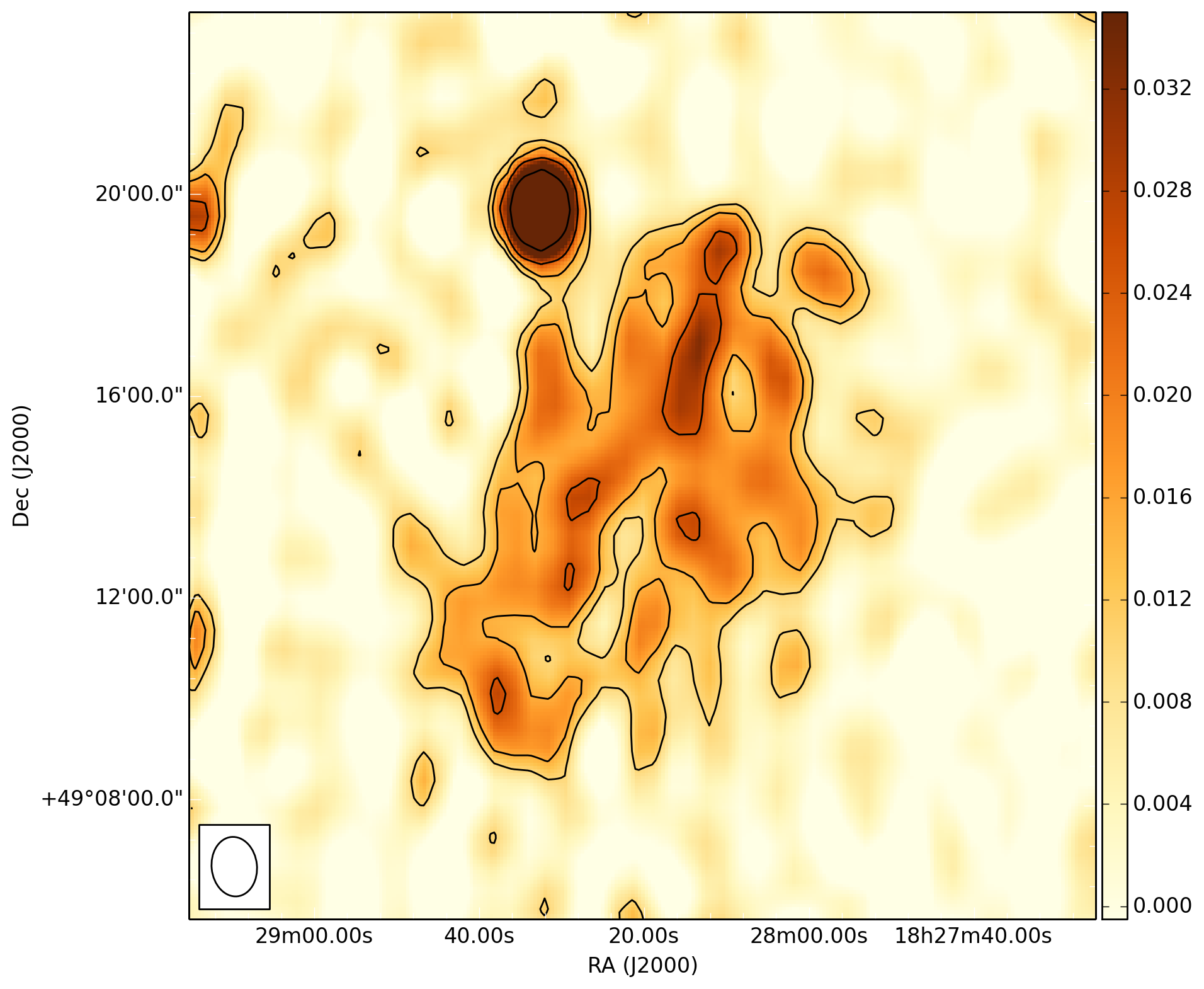}
        
\centering
\includegraphics[width=0.4\textwidth]{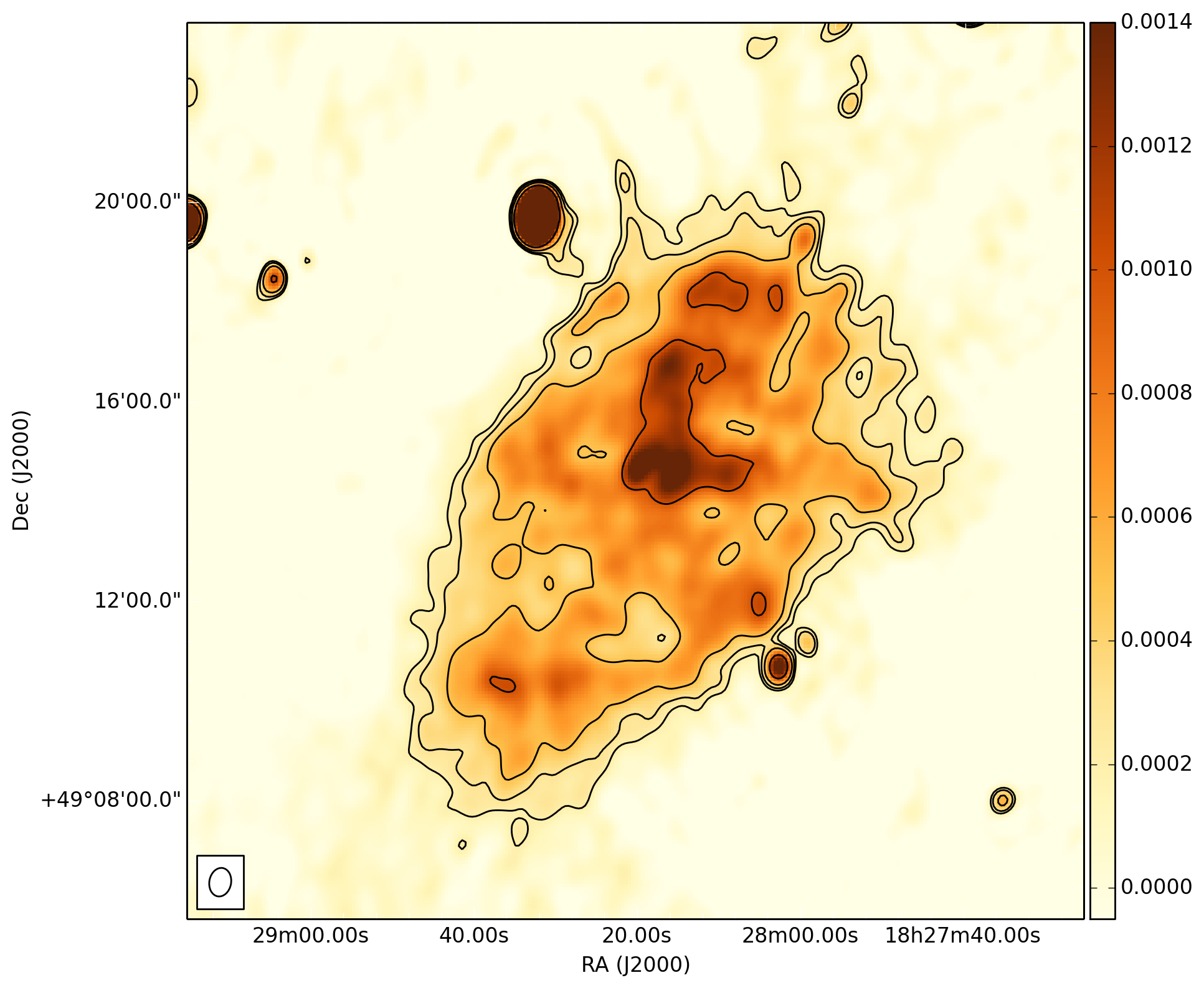}
\caption{Radio maps of blob1: (top panel) LOFAR 137-MHz map, levels 2, 3, 5, 10 $\times \ \sigma \ (3.5 \ \rm mJy \ beam^{-1})$; (middle panel) WENSS 327-MHz map, levels 2, 3, 5, 10 $\times \ \sigma \ (5 \rm \ mJy \ beam^{-1})$; (bottom panel) WSRT 1400-MHz map, levels: 2, 3, 5, 10 $\times \ \sigma \ (0.08 \ \rm mJy \ beam^{-1}) $.}
\label{fig:maps}        
        
\end{figure}

The radio images obtained from the data described in Sect. \ref{data} are presented in Fig.~\ref{fig:maps}. At all available frequencies, blob1 shows a very diffuse amorphous shape emission with low surface brightness with a mean value of 4 mJy $\rm arcmin^{-2}$ at 1.4 GHz. Its size is $\rm 12\arcmin\times7\arcmin$, corresponding to an axial ratio of 1.7. We do not find evidence for edge brightening or sharp features such as jets or hot spots, but there is a broad linear enhancement of surface brightness extending to the north in both the 1.4-GHz WSRT and the LOFAR map. In the high-resolution 1.4-GHz WSRT image, which has a resolution of 15", we also identify a compact component with a flux density $\rm S_{1.4GHz}=1$ mJy located near the centre of the diffuse emission. This feature is also confirmed by the 4.9-GHz WSRT observation with $\rm S_{4.9GHz}=0.7$ mJy (see Fig.~\ref{fig:counterpart}).

\begin{small}
\begin{table}[t]
        \centering
        \caption{List of two-point spectral indices and respective errors as a function of frequency. }
        \small
                \begin{tabular}{l l}
                \hline
                \hline
                Frequency (MHz)& Spectral index\\
                \hline
                116-155 & $0.50\pm1.0$ \\
                155-327 & $0.55\pm0.40$ \\
                327-1400 & $0.90\pm0.20$  \\            
       1400-4850 & $\geq$ 1.60\\
                \hline
                \hline  
                \end{tabular}
                \label{tab:spec_index}
\end{table}
\end{small}

\subsection{Optical identification and galactic environment}
\label{opticalidentifcation}

\begin{figure*}[h!]
\centering
\includegraphics[width=1.2\textwidth]{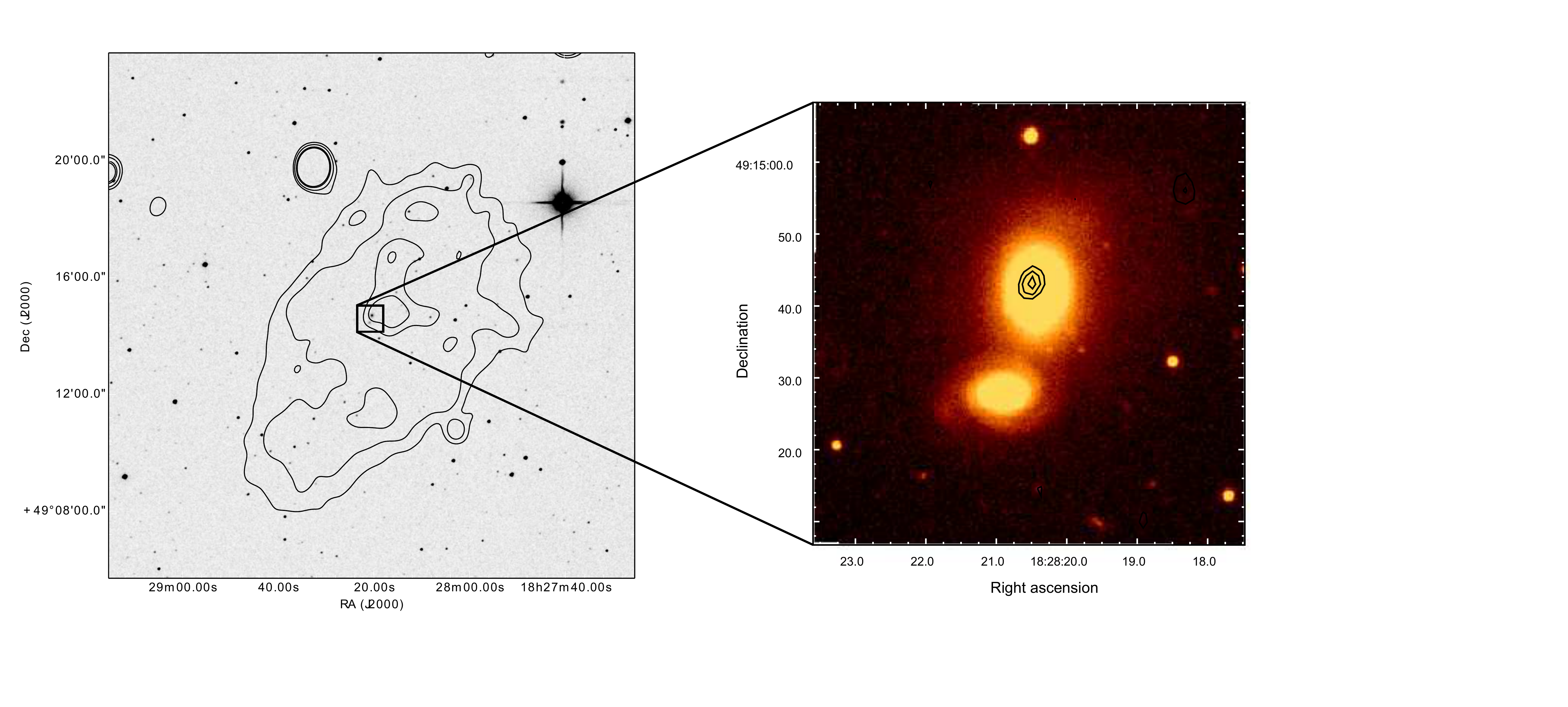}
        \caption{\textit{Left - }blob1 contours from the 1.4 GHz WSRT map (levels 3, 6, 12, and 15 $\sigma$) overlaid on the 2MASS $\rm K_s-band$ image of the field; \textit{right -} Zoom-in on the host galaxy and on the companion galaxy from the $r$-band image contours from the 5 GHz WSRT map in black (levels 3, 5,
and 7 $\times \ \sigma$ \ ($0.085 \ \rm mJy \ beam^{-1}$).}
\label{fig:counterpart} 
        
\end{figure*}

Using data from the Digital Sky Survey (DSS), we identify two galaxies located near the centre of the diffuse radio source;
these are the brightest optical objects within the field (see Fig. \ref{fig:counterpart}). The brightest of this pair of galaxies is co-located with the compact component detected in the WSRT images described in Sect. \ref{morphology} and is considered the most likely host. The galaxies have been observed as part of the Two Micron All-Sky Survey (2MASS) and are catalogued as 2MASXJ18282048+4914428 (1) and 2MASXJ18282089+4914278 (2). They have $\rm m_1$= 12.9 and $\rm m_2$=13.8 in the $\rm K_s-band$ (Vega) at 2.17 $\rm \mu m,$ respectively.

We obtained optical spectra for the two galaxies on 25 February 2015 and deep imaging in a filter equivalent to the $r$ band on 30 April 2015 with the ISIS spectrograph and the ACAM instrument, both mounted on the 4.2 m William Herschel Telescope (WHT). The spectra were reduced using standard  \texttt{IRAF}\footnote{IRAF is distributed by the National Optical Astronomical Observatories, which are operated by the Association of Universities for Research in Astronomy, Inc., under cooperative agreement with the National Science Foundation} tasks. The redshifts of the galaxies are $z_1 = 0.051$ and $z_2 = 0.052$, both with relative uncertainties of $\rm \Delta z = 0.00025$, and absolute uncertainties of $\rm \Delta z = 0.001$. These are consistent with the photometric redshifts from the 2MASS Photometric Redshift catalogue  \citep[2MPR][]{bilicki2014} of $\rm z_{phot,1} = 0.053\pm0.006$ and $ \rm z_{phot,2}=0.050\pm0.006$.

The spectra of the two galaxies appear to be typical of early-type galaxies with strong 4000\AA \ breaks and stellar absorption features. No strong emission lines are detected in the spectra that might be connected to an optical AGN or to star formation. An upper limit on the star formation rate can be derived from the [OII] emission line flux, using the reddening-corrected relation between [OII] luminosity,  L[OII], and star formation rate derived in Eq. 4 from \cite{kewley2004}. A synthetic unresolved [OII] 3727,3730\AA \ doublet of width 3\AA \ and peak flux equal to 3-$\sigma$ of the noise measured between 3800--4100\AA \ (3615--3900\AA \ rest-frame) was inserted into the spectra of the two galaxies. The star formation rates resulting from this emission line are $\rm SFR_1$ = 0.51 and $\rm SFR_2$ = 0.15 $\rm M_{\odot}/year$. Unfortunately, $\rm H\alpha$ falls directly on a telluric absorption line, which
means that no information can be derived from this portion of the spectrum. The $\rm H\beta-[OIII]$ region in either spectrum shows no signs of AGN activity.

If the radio emission is associated with the two galaxies, it
would imply a linear size for the radio emission of $\rm 700 \times 410$ kpc. In Fig. ~\ref{fig:counterpart} the 1.4-GHz WSRT radio contours are overlaid on the 2MASS map, showing the extent of the radio emission. From the $\rm K_s-band$ luminosity and assuming a mass-to-light ratio of 0.6 \citep{mcgaugh2014}, we derived a first-order stellar mass estimate of $\rm logM_1=10.6 \ M_{\odot} $ and $\rm logM_2=10.3 \ M_{\odot}$.

The close physical separation and radial velocity difference of $\rm \sim 300 \ km \ s^{-1}$ suggest that the galaxies are interacting. This is confirmed by tidal features observed in the $r$-band image, as is shown in Fig. \ref{fig:counterpart} and further discussed in Sect. \ref{mergerscenario}. 

The two galaxies do not belong to any known galaxy cluster or group. The closest galaxy cluster is MCXC J1811.0+4954 (z=0.0501), which is located at a distance of 2.7$^{\circ}$ $\simeq$ 10 Mpc \citep{bohringer2000}. In our deep r-band image we identify at most three or four candidate members within a projected radius of 50 kpc from the host galaxy. No spectral information is available for these objects, therefore we cannot verify if they are background or foreground sources. We therefore conclude that the pair of galaxies might be part of a small group, but probably not part of a massive cluster.

\subsection{Source energetics}
\label{sourceenergetics}
 
In this and the following section, we derive physical parameters of the radio emission and model the radio spectrum to assess the nature of blob1 and how it compares with other sources. Assuming J18282048+4914428 to be the host galaxy, the radio luminosity of the diffuse emission at 1.4 GHz is $\rm L_{\rm 1.4 GHz} = 1.5\times 10^{24} \rm \ W Hz^{-1}$.

We computed the magnetic field strength to be 1 $\rm \mu$G and the total energy density to be $1.6 \times 10^{-13}$ $\rm erg \ cm^{-3}$ , assuming the plasma to be in equipartition condition between particles and the magnetic field using the derivation of \cite{beck2005}. This implies a non-thermal pressure of $5 \times 10^{-14}$ $\rm dyne \ cm^{-2}$. For this calculation we assumed a power-law particle distribution of the form $\rm N(\gamma) \propto \gamma^{-a}$ between a minimum and maximum Lorentz factor of $\rm \gamma_{\rm min} = 10$ and $\gamma_{\rm max} = 10^6$. We calculated the volume of the source as $\rm 1.7\times10^{72} \ cm^3$ assuming a prolate ellipsoidal geometry with the major and minor axis equal to the major and minor projected size of the source. We furthermore assumed the particle energy content of the source to be equally distributed between heavy particles and electrons so that their ratio k=1, and we set $\alpha=0.5$ according to the observed low-frequency spectral shape where radiative energy losses are less significant. 
We note that with this low value for the magnetic field, the radiative cooling of the plasma is dominated by inverse Compton (IC) scattering of cosmic microwave background (CMB) photons: the energy density of the CMB is a factor 10 greater than the magnetic energy density.

\subsection{Spectral properties and age}
\label{spectralpropertiesandages}

The spectral shape of radio sources is directly related to the electron populations responsible for the emission and gives important constraints about their age, as we describe below \citep{kellerman1964, pacholczyc1970}.
The integrated radio spectrum of active galaxies is well described by the continuous injection model \citep[CI,][]{kardashev1962}, which assumes a continuous replenishment of new relativistic particles with an energy distribution such that $\rm N(E)\propto E^{-p}$.  This results in a broken power-law radiation spectrum with spectral index $\rm \alpha_{inj}=\frac{p-1}{2}$ below a critical frequency $\rm \nu_{b, low}$ and $\rm \alpha = \alpha_{inj}+ 0.5$ above $\rm \nu_{b, low}$ that is due to preferential radiative cooling of high-energy particles. The typical $\rm \alpha_{inj}$ of active sources is in the range of 0.5-0.8 and, as such, has spectral index $\alpha =1 \sim 1.3$ at high frequency. When the nuclear activity switches off, a new break frequency $\rm \nu_{b,high}$ appears beyond which the spectrum drops exponentially \citep{komissarov1994}. Knowledge of the spectral shape and the source energetics allows estimating the age of the electron population according to radiative cooling as follows:

\begin{equation}
t_s=1590\frac{B_{\rm eq}^{\rm 0.5}}{(B_{\rm eq}^2+B_{\rm CMB}^2)\sqrt{\nu_{\rm b}(1+z)}} \\ ,
\label{eqtime}
\end{equation}

\

\noindent where $t_s$ is given in Myr, the magnetic field $B$ and inverse Compton equivalent field $B_{CMB}$ are in $\rm \mu G,$ and the break frequency $\rm \nu_{\rm b}$ is given in GHz. Furthermore, the ratio between  $\rm \nu_{b, low}$  and  $\rm \nu_{b, high}$ constrains the active time of the source with respect to the total age as follows:

\begin{equation}
\frac{\rm t_s}{\rm t_{OFF}}=\left( \frac{\rm \nu_{b,high}}{\rm \nu_{b,low}}\right)^{0.5} ,
\label{eqage}
\end{equation}

\

\noindent where $\rm t_s$ is the total age of the source and can be written as $\rm t_s=t_{\rm CI}+t_{\rm OFF}$ , where $\rm t_{CI}$ is the continuous injection phase and $\rm t_{OFF}$ is the off phase.

\

\begin{figure}[h]
\centering
\includegraphics[width=0.5\textwidth]{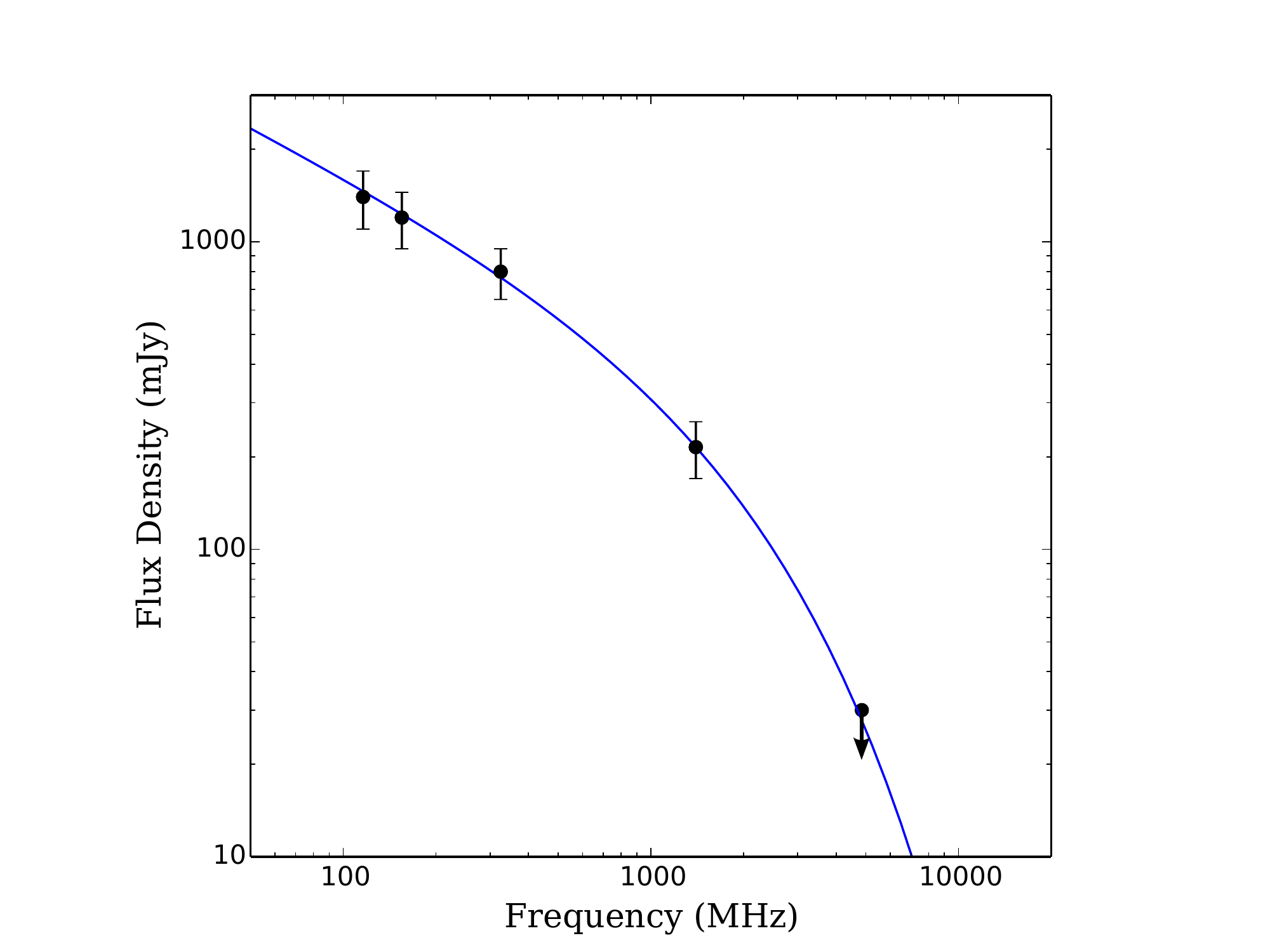}
        \caption{Integrated radio spectrum of blob1. The solid line represents the best fit of the synchrotron model described in Sect. \ref{spectralpropertiesandages}. The flux densities used for the fitting are summarized in Table \ref{tab:fluxes}.}
\label{fig:spec}
\end{figure}

As shown in Fig.~\ref{fig:spec}, the spectrum of blob1 is highly curved at a frequency $\geq 1.4$ GHz. The measured spectral indices at low frequency are in accordance with typical observations of active radio sources (see Table~\ref{tab:spec_index}). At higher frequencies, the spectral shape steepens to reach a value of $\rm \alpha_{1400}^{4850}\geq1.6$. Following \cite{murgia2011}, we computed the 
spectral curvature $\rm SPC=\alpha_{high}-\alpha_{low}$ using $\rm \alpha_{low}=\alpha_{116}^{327}$ and $\rm \alpha_{high}=\alpha_{1400}^{4850}$ and obtained a value of $\rm SPC\geq1.1$. This spectral curvature is well beyond the values expected for active radio sources according to the continuous injection model discussed above (SPC $\leq$ 0.5), therefore it suggests that the source of the particle injection is switched off.

Because the slope variation of the spectrum is so abrupt, we can assert that the two break frequencies  $\rm \nu_{b, low}$  and $\rm \nu_{b,high}$ lie in this frequency range and are close to each other. This implies, according to Eq. \ref{eqage}, that the source has been inactive for a significant fraction of its total age. Assuming  $\rm \nu_{low}\geq1.4$ GHz and $\rm \nu_{high}\leq5$ GHz, we can calculate to first order the total age and the off phase of blob1 using Eq.\ref{eqtime} for synchrotron cooling. From this we derive the following limits for the source total age and off phase: $\rm t_s\leq95$ Myr and $\rm t_{OFF}\geq45$ Myr.

To obtain a more accurate estimate of the activity timescales of the source, we fit the spectrum with a model describing the spectral evolution of a source that is active for a time $\rm t_{\rm CI}$ and then switches off for a time $\rm t_{\rm OFF} = t_s- t_{CI}$ \citep{komissarov1994, murgia2011}. For the modelling we assumed a fixed injection index of $ \rm \alpha_{inj} =0.5$ and the magnetic field value 1 $\rm \mu G$ derived in Sect. \ref{sourceenergetics}. The best fit is shown in Fig.~\ref{fig:spec} and the best-fit parameters are $\rm t_{\rm OFF} = 60^{+15}_{-20} $ Myrs and $\rm t_s/t_{OFF}=1.3^{+1.7}_{-0.3}$ (see Table \ref{tab:spec_fitting}). This implies an active phase of $\rm t_{\rm CI} = 15^{+65}_{-14}$ Myr and a total age of $\rm t_s=75^{+50}_{-10}$ Myr, which agrees with the previous first-order estimates. We note that because of the assumptions and systematic errors that enter the equation, these numbers should only be considered as characteristic timescales. 

\begin{small}
\begin{table}[t]
        \centering
        \small
        \caption{Best-fit parameters of the spectral modelling presented in Sect. \ref{spectralpropertiesandages}.}
                \begin{tabular}{l l l l l l l}
                \hline
                \hline
                \rm $\alpha_{\rm inj}$ & B & $\rm t_s/t_{OFF}$ &  $t_{\rm OFF}$ \\
                & $\mu G$  &  & Myr  \\ 
                \hline
                \\
                0.5 fixed & 1 & $1.3^{+1.7}_{-0.3}$  &  $60^{+15}_{-20}$\\
                \\
                \hline
                \hline  
                \end{tabular}
                \label{tab:spec_fitting}
\end{table}
\end{small}

\section{Discussion}
\label{discussion}

\subsection{Evidence in favour of the remnant radio galaxy interpretation}
\label{phenomenologyofblob1}

The morphology of blob1 with its amorphous shape and low-surface brightness together with its spectral curvature supports an interpretation in which the source is a remnant of a past radio AGN activity. Here we discuss these aspects in detail in the context of current knowledge of the population of active and remnant radio galaxies.  

According to the P-D (power-size) diagram for low-power radio galaxies by \cite {parma1999}, blob1 has a relatively large size for its luminosity. This may suggest that the central engine was more powerful during the maximum of its activity. As the jets switch off, powerful radio galaxies are expected to undergo a rapid luminosity decrease \citep{slee2001, kaiser2002, hardcastle2013}. Another explanation could be that the radio source has expanded in an environment with a comparatively low density.  However, blob1 still lies within the broad scatter of the P-D correlation,
therefore this is not conclusive. 

The core of blob1 observed at 1.4 GHz and 4.9 GHz has a power of $\rm 5.7 \times 10^{21} \ W \ Hz^{-1}$ at 1.4 GHz, giving a core prominence (i.e. ratio between the radio power of the core at 5 GHz and the extended emission at 178 MHz) of about $6\times10^{-4}$. This value lies more than a factor 10 below the correlation observed for active radio galaxies by \cite{giovannini1988}. This suggests that the nuclear engine may be still active in a low-power state and that the broad linear feature presented in Sect. \ref{morphology} might be related to an uncollimated weak outflow connected to the core. The spectral index of the core is $\rm \alpha_{1400,core}^{4900}=0.3$ in accordance with its AGN nature. Star formation is unlikely since the 1.4 GHz core luminosity would translate into a star formation rate of 1.2 $\rm M_{\odot} \ yr^{-1}$ \citep{condon2002}, more than twice
as much as what is derived from the optical spectrum. The lack of strong AGN emission lines is also consistent with the central engine being in a dying phase or in a low-power 'hot mode' state \citep{tadhunter1998}.

Remnant radio galaxies are expected to lack compact features typical of active galaxies, such as defined jets or hot spots. Depending on whether the central activity has ceased gradually or abruptly, they can also show weak or absent cores. The morphology and evolution of the remnant does depend on the external medium and on the properties of the progenitor active source, which are different for FRI and FRII radio galaxies. \cite{saripalli2012} suggested the following criteria for distinguishing FRI from FRII remnant sources on a morphological basis: i) FRII remnant sources appear as pairs of diffuse edge-brightened lobes lacking hot spots, jets, and cores; ii) FRI remnant sources do not show any compact emission except for a weak core with twin extensions or trails that do not resemble typical jets or lobes. With its relaxed lobes (axial ratio of 1.7), absence of edge brightening, and weak core, blob1 is closer to an FRI morphology. According to \cite{parma1996}, about 62\% of the B2 sample of low-luminosity radio galaxies have double-lobe morphology with relaxed lobes whose typical axial ratio is 2, whereas high-power FRII radio galaxies show much higher axial ratios of about 5. The standard prototypes of the ``fat double'' sources (Owen \& Laing 1989) are identified with 3C310; 3C314.1 and 3C386 having power in the range $\rm S_{178}=10^{24}-10^{25} \ W \ Hz^{-1} sr^{-1}$ and axial ratios $\leq2$. Hydrodynamical simulations by \cite{stenberg2007} showed that such fat bubbles of plasma can be inflated by wide and slow jets. It is worthwhile mentioning that \cite{subrahmanyan2006} also claimed that the relaxed or extended morphology observed in some sources might be the result of jet precession during the fly-by encounters that precede a merger event. In this case, the synchrotron plasma is deposited over a wide range of angles conferring the observed morphology. As discussed in Sect. \ref{opticalidentifcation}, the host galaxy of blob1 is interacting with a companion, therefore this last scenario may also be relevant to the evolutionary history of blob1.

\subsection{Merger scenario and triggering mechanism}
\label{mergerscenario}

A complete investigation of the evolutionary history of blob1 involves the analysis of the mechanism responsible for triggering and stopping the radio-loud AGN. Because the host galaxy of blob1 is currently interacting with another galaxy, we should consider the possibility that this process had a role in marking the timescales of the radio activity of the AGN.

Galaxy mergers are often suggested to influence the triggering and quenching of the jets in powerful radio-loud AGN \citep{tadhunter2012, wen2012}. They allow the gas to lose angular momentum and flow towards the nuclear region of the galaxy, whereby they either contribute to replenishing the black hole fuel, which in turn
enhances the jet production, or to unsettling the accretion, which causes the disruption of the jets. Tidal encounters between galaxies that precede the final coalescence can also be responsible for triggering or stopping the AGN activity \citep{ellison2011}. In this phase, intermittent jets can be produced, and if the accretion disk is perturbed, they can also be realigned. One of the best-studied example of this occurrence is CenA, where a merger is thought to be responsible for the interruption of the radio activity for a time of the order of 10 Myr, after which new-born jets have restarted \citep{saxton2001, struve2010}. Although statistical evidence of the link between mergers and radio emission has been found, the AGN radio activity has not been observed to be associated to a specific phase of the interaction \citep{ramosalmeida2011}.

From the $r$-band image it is clear that the morphology of the two galaxies is disturbed. The host galaxy shows an asymmetry in the brightness distribution on the top east side while the companion shows clear tidal features on the west and east side. Although mergers with late-type galaxies can produce the most prominent tidal features, early-type galaxies are also observed to produce tidal tails \citep{ramosalmeida2011}, but much broader, diffuse, and shorter lived than what is observed in gas-rich
mergers. Since both galaxies are early-type, we do not expect much gas to be involved in the interaction and therefore not much star formation to be going on. This agrees with the computed star formation rate presented in Sect. \ref{opticalidentifcation}. 
According to their morphology, it is likely that the two galaxies are observed after the first pericentre passage but before the final coalescence of the merging nuclei.

We estimate the dynamical timescale of the system $\rm t_{dyn}=60 \ Myr$ based on the ratio between the projected separation between the two galaxies $\rm d_{proj}=15 \ kpc$ and their relative speed $\rm v_{rel}=300 \  km \ s^{-1}$. Although this timescale is uncertain and needs to be taken with care, its similarity to the $\rm t_{OFF}$ of the radio source may suggest that the interaction could have disrupted the radio jets.

\subsection{Spectral comparison with other remnant radio galaxies and implications for remnant evolution}
\label{spectralcomparison}

Based on the discussion in Sect. \ref{phenomenologyofblob1}, we consider blob1 to be a remnant radio galaxy and compare here its spectral characteristics with those observed in previously selected sources of the same class. In particular, we mainly focus our comparison on the only complete samples of dying radio galaxies reported by \citet{parma2007} and \citet{murgia2011}, for which a uniform multi-frequency analysis is available. \citet{parma2007} selected their sample using $\rm \alpha_{327}^{1400} >1.3,$ while \citet{murgia2011} selected their sample based on the spectral curvature defined in Sect.\ref{spectralpropertiesandages}. 

Although it is generally believed that remnant radio galaxies develop spectral indices > 1 below 1.4 GHz, this is not the rule, and blob1 is one of the few known examples of remnant sources with low-frequency spectra consistent with active sources. A similar case is the source B2 1610+29 in \citet{murgia2011}. These remnant galaxies show the high-frequency spectral cut-of
that is typical of the dying phase only above 1 GHz, while at lower frequency they still preserve the injection spectral index that characterizes the active phase. This spectral shape can either be justified by the plasma being still young or by a slow rate of radiative cooling. 
 
The spectral modelling of blob1 presented in Sect. \ref{spectralpropertiesandages} shows that the source has been spending about 50-80\% of its life in a dying phase. The computed total age agrees with the mean value observed by \citet{murgia2011} and \citet{parma2007} and also with what has been derived in other individual sources such as B2 0924+30 \citep[60 Myr;][Shulevski et al. in prep]{jamrozy2004}, B 0917+75 \citep[100 Myr;][]{harris1993}, J1324-3138 \citep[97 Myr;][]{venturi1998}, and A 2622 \citep[150 Myr;][]{giacintucci2007}. The same can be asserted for the ratio $\rm t_{OFF}/t_{s} $ ,
which fits the range of values 0.2-0.8 computed in \citet{murgia2011} and \citet{parma2007} within the errors. The difference in spectral shape can thus be explained by different values for magnetic fields (sources in \cite{murgia2011} have $\rm B_{eq}= 4-45 \ \mu G$ compared with 1 $\rm \mu  G$ in blob1). 

It is interesting to note that despite being associated with a field galaxy, blob1 has a long dying phase whose duration is comparable to those observed for remnants in clusters. Dense environments are thought to keep the plasma confined, preventing adiabatic expansion and allowing their detection for a longer time. Instead, in low-density environments, the plasma is free to expand, and if it is over-pressured with respect to the external medium, the source can rapidly fade away. This argument is often invoked to explain why so few remnants are detected outside clusters, although a difference in the duty cycle of radio galaxies in different environments can also be an explanation \citep[e.g.][]{murgia2011}. To date, very few other remnant sources outside clusters have been detected and studied: 0924+30 \citep{jamrozy2004}, NGC 5580 \citep{degasperin2014}, and NGC 1534 \citep{hurleywalker2015}. 

The reason why the plasma is still visible after such a long time has to be connected with its energetics. If the lobes have low internal energy densities and pressures, their expansion rate is low and the radiative cooling is the dominant process in the evolution of the source.
This is very likely the case for sources with a low surface brightness like blob1 \citep{subrahmanyan2003}. Furthermore, it has been argued that most sources will be at pressure equilibrium at the end of their lives \citep{hardcastle2013}, but without further information on the external medium we cannot evaluate whether the plasma is over-pressured or in pressure balance. The only archival X-ray observations of this area of sky are from the ROSAT survey, and because they are too shallow, they do not yield useful constraints.  

Better comprehension of the phenomena that drive the plasma evolution as a function of energy and environment can only be achieved with larger samples of this class of sources. For this reason new systematic searches exploiting new-generation instruments such as LOFAR need to be performed, as we discuss in more detail in the following section.

\section{Implications for selection of remnant radio galaxies}
\label{implicationsforsourceselection}

Until now, systematic searches of remnant radio galaxies have mostly been carried out by cross-matching radio surveys at different frequencies using the steep spectral index as source selection criterion (typically $\alpha > 1.3 $). \cite{parma2007} selected a sample of dying radio galaxies from 327-MHz WENSS and the 1.4-GHz NVSS catalogues. \cite{dwarakanath2009} and \cite{vanweeren2009} extracted a sample of remnant candidates from the 74-MHz VLSS and 1400-MHz NVSS surveys. A few tentative searches have also
been performed on individual deep fields \citep{sirothia2009, vanweeren2014}. The strategy of the steep spectral index is best for identifying sources with very curved spectra at low frequencies that have experienced significant radiative losses and are typically not detectable at 1.4 GHz. 

The remnant radio galaxy presented in this paper demonstrates that depending on its physical conditions and environment, the plasma can undergo a different evolutionary path and present different spectral shapes. Therefore a number of sources  might be missed by this standard selection approach and new selection methods should be considered. \cite{murgia2011} suggested that the SPC can be a powerful diagnostic tool for tracing the evolutionary stage of radio sources. On the basis of synchrotron models, dying sources are expected to have an SPC > 0.5. With this method, they selected a sample of five dying radio galaxies by combining multi-frequency survey data and dedicated observations. The SPC measured for blob1 supports the reliability of the strategy. The broad range of frequencies required to compute it makes it hard to apply it systematically as selection criterion, however. The lack of deep, all-sky, high-frequency surveys ($\sim$5 GHz) represent a limitation for this kind of approach. A combination of low frequency (e.g. 60-600 MHz) and high frequency (e.g. 1400-8000 MHz) is indeed needed to properly trace the spectral curvature \citep{harwood2013}. We also note that sources with power-law non-steep spectra like the one studied by \cite{degasperin2014} would likewise be missed with this method.

In light of this, we suggest that the population of remnant radio galaxies cannot be fully characterized by a single selection criterion based on spectral index or spectral curvature, for
instance, but these should rather be used in a complementary way. Given the variety of observed spectral shapes and to construct complete samples, morphological selection should also play a role alongside spectral criteria. Visual inspection of radio sources has already provided a number of detections in addition to the source presented here \citep[e.g.][]{degasperin2014, harris1993, hurleywalker2015, giovannini1991}, but it still has not been applied in a systematic way. Tentative searches based on morphological characteristics have only been performed by \cite{saripalli2012} in the context of the Australia Telescope Low Brightness Survey and by \cite{jones2001} using the VLSS. Because remnant sources are expected to be very diffuse with very low surface brightness emission, new-generation instruments such as LOFAR with unprecedented sensitivity capabilities are expected to unveil a new population of such sources. The mean surface brightness of blob1 (equal to 4 $\rm  mJy \ arcmin^{-2}$ at 1.4 GHz) is indeed among the lowest values ever observed, and it lies just at the border of what can be detected with currently available surveys. This value is only comparable to similar sources such as the remnant PKS B1400-33 \citep[3 $\rm  mJy \ arcmin^{-2}$,][]{subrahmanyan2003}, the remnant B2 0258+35  \citep[1.5 $\rm  mJy \ arcmin^{-2}$][]{shulevski2012}, several restarted FR-I and FR-II radio galaxies detected in the Australia Telescope Low Brightness Survey \citep[2.5-6 $\rm  mJy \ arcmin^{-2}$,][]{saripalli2012}, and a few active giant radio galaxies such as J0034.0-6639 \citep[1 $\rm mJy \ arcmin^{-2}$,][]{saripalli2012} and SGRS J0515-8100 \citep[10 $\rm mJy \ arcmin^{-2}$,][]{subrahmanyan2006}.

In conclusion, deep imaging at low frequency will surely probe a fainter population of steep-spectrum sources, which will expand the previous samples obtained from VLSS, WENSS, and NVSS, but will most likely also reveal diffuse sources with a flatter spectrum, like blob1. To avoid selection biases towards steep integrated radio spectra, additional selection criteria should play a role in identifying remnant AGN candidates; these would be spectral curvature and morphology. A complete representation of the remnant radio galaxy population can be provided only in
this way. By increasing the detection of these rare sources, we will finally be able to perform uniform systematic studies that will improve our knowledge of the different evolution histories of radio galaxies and of the different mechanisms that drive the remnant plasma evolution after the jets switch off.

\begin{acknowledgements}
We thank Aarthi Ramesh, Simon Steendam, Reynier F. Peletier and S. Trager for performing the WHT observations on April 30, 2015 and for the useful comments. LOFAR, the Low Frequency Array designed and constructed by ASTRON (Netherlands Institute for Radio Astronomy), has facilities in several countries that are owned by various parties (each with their own funding sources) and that are collectively operated by the International LOFAR Telescope (ILT) foundation under a joint scientific policy. The Westerbork Synthesis Radio Telescope is operated by ASTRON with support from the Netherlands Foundation for Scientific Research (NWO).
The research leading to these results has received funding from the European Research Council under the European Union's Seventh Framework Programme (FP/2007-2013) / ERC Advanced Grant RADIOLIFE-320745. This research has made use of the NASA/IPAC Extragalactic Database (NED),
which is operated by the Jet Propulsion Laboratory, California Institute of Technology,
under contract with the National Aeronautics and Space Administration. This research made use of APLpy, an open-source plotting package for Python hosted at http://aplpy.github.com. This publication makes use of data products from the Two Micron All Sky Survey, which is a joint project of the University of Massachusetts and the Infrared Processing and Analysis Center/California Institute of Technology, funded by the National Aeronautics and Space Administration and the National Science Foundation.
      
\end{acknowledgements}

\bibliographystyle{aa}
\bibliography{BLOB1_Brienza.bib}

\end{document}